\documentclass[prb,aps]{revtex4}
\usepackage{graphicx}
\usepackage{amsmath}
\usepackage{bm}
\def\qr{{\bf r}}                                   

\newcommand{\e}{{\rm e}}

\newcommand{\epp}{\epsilon}
\newcommand{\ep}{\epsilon}

\newcommand{\be}{\begin{equation}}
\newcommand{\ee}{\end{equation}}
\newcommand{\ba}{\begin{eqnarray}}
\newcommand{\ea}{\end{eqnarray}}

\newcommand{\nn}{\nonumber}
\newcommand{\la}{\label}

\begin{document}

\title{Extrapolated High-Order Propagators for Path Integral Monte Carlo Simulations}

\author{Robert E. Zillich}
\affiliation{Institut f\"ur Theoretische Physik, Johannes Kepler
Universit\"at Linz, A-4040 Linz, Austria}
\author{Johannes M. Mayrhofer}
\altaffiliation[present address: ]{Institut f\"ur Pharmakologie und Toxikologie,
Universit\"at Z\"urich, Switzerland}
\affiliation{Institut f\"ur Theoretische Physik, Johannes Kepler
Universit\"at Linz, A-4040 Linz, Austria}
\author{Siu A. Chin}
\affiliation{Institut f\"ur Theoretische Physik, Johannes Kepler
Universit\"at Linz, A-4040 Linz, Austria}
\affiliation{Department of Physics,
Texas A\&M University, College Station, TX 77843, USA}

\begin{abstract}
We present a new class of high-order imaginary time propagators for
path-integral Monte Carlo simulations by subtracting
lower order propagators. By requiring all terms of the extrapolated propagator
be sampled uniformly, the subtraction only affects the potential part of the path
integral.  The negligible violation of positivity of the resulting path integral 
at small time steps has no discernable affect on the accuracy of our method.
Thus in principle arbitrarily high order
algorithms can be devised for path-integral Monte Carlo simulations.
We verify this claim is by showing that fourth, sixth, and eighth order convergence
can indeed be achieved in solving for the ground state of strongly interacting
quantum many-body systems such as bulk liquid $^4$He.
\end{abstract}


\maketitle


\section{Introduction}

Many quantum Monte Carlo (QMC) techniques, such as path integral [ground state] Monte Carlo
(PI[GS]MC) and diffusion Monte Carlo (DMC),
rely on stochastic propagation of the Schr\"odigner equation in imaginary time.
In all these methods, the probability distribution sampled is the matrix element,
or the trace, of the imaginary time propagator
\be
  G(\tau) = \e^{-\tau H} = \e^{-\tau (T+V)},
\ee
with Hamiltonian $H=T+V$ and kinetic and potential operators
$T=(-\hbar^2/2m)\sum_{i} \nabla^2_i$ and $V=\sum_{i<j}v(r_{ij})$.
Since $G(\tau)$ is generally unknown, $\tau$ is usually
discretized into a sum of short time steps $\ep$ so that
the full propagator $G(\ep)$ can be approximated by
a product of short-time approximate propagator $\tilde G(\ep)$.
The computational effort, and therefore the efficiency of these QMC techniques depends
on $\ep$. If $\tilde G(\ep)$ is accurate to
high orders in $\ep$, then a large $\ep$ can be
used to span a given imaginary time interval, resulting in fewer
samplings of (but possibly more complex) $\tilde G(\ep)$.

For QMC simulations, there is a surprise lack of
general higher order algorithms. For example, one has the
well known second-order, primitive propagator
\be
  G_2(\ep)=\e^{-\ep V/2}\e^{-\ep T}\e^{-\ep V/2}
  = G(\ep) + {\rm O}(\ep^3).
\la{g2form}
\ee
(For computing the trace, any splitting first-order algorithm,
such as $\e^{-\ep V }\e^{-\ep T}$,
will also yield a second-order trace\cite{chin04}.)
The highly successful pair density propagator,
that approximates $G$ by a pair-wise product of exact two-body
propagators~\cite{ceperleyRMP95} must also be second order in
the general many-particle case, but possibly with a small error coefficient.
However, this approach
in practice is limited to solely spherically symmetric
interactions due to the difficulty of evaluating the
two-particle density matrix exactly. The only fourth-order
method known for many years is the Takahashi-Imada\cite{ti},
Li-Broughton\cite{li} propagator
\be
G_{\rm TI}(\ep)=\e^{-\ep T/2}e^{-\ep V-(\ep^3/24) [V,[T,V]]}e^{-\ep T/2}
= G(\ep) + {\rm O}(\ep^3)\,,
\ee
where $[V,[T,V]]=\frac{\hbar^2}m\sum_i|\nabla_i V|^2$.
This ``corrector'' propagator is only second order, but yields
a fourth-order trace, as explained in Ref.\onlinecite{chin04}. Thus until
recently, there were only two second-order and one fourth-order
algorithm for PIMC simulations.

The problem of constructing higher order PIMC algorithms is the
{\it time-irreversible} nature of the imaginary time Schr\"odinger
equation. The short-time propagator can in general be approximated
to any order by a product decomposition,
\be
{\rm e}^{-\ep (T+V )}\approx\prod_{i=1}^N
{\rm e}^{-t_i\ep T}{\rm e}^{-v_i\ep V},
\label{arb}
\ee
with coefficients $\{t_i, v_i\}$ determined by the required order of accuracy.
However, in QMC applications, since
$\langle R^\prime|\,{\rm e}^{-t_i\ep T}|\,R\rangle
\propto {\rm e}^{-(R^\prime-R)^2/(4 Dt_i\ep)}$
is the diffusion kernel with $D={\hbar^2/2m}$,
the coefficient $t_i$ must be
positive in order for the kernel to be normalizable as a
probability distribution. As first proved by Sheng\cite{sheng} and
Suzuki\cite{suzukinogo}, and later by Goldman-Kaper\cite{goldman}
and Chin\cite{chin063},
beyond second order, any factorization of the form (\ref{arb})
{\it must} contain some negative
coefficients in the set $\{t_i, v_i\}$.
Thus, despite myriad of higher-order propagators of the single product
form (\ref{arb}) for solving the {\it time-reversible}, real-time Schr\"odinger equation,
none can be applied in PIMC beyond second order. It is only in the last
decade that bona fide fourth order, {\it forward} algorithms
with all positive coefficients have been found\cite{suzfour,chin97}
and applied to DMC and PIMC simulations\cite{fchinl,fchinm,jang}.
In order to circumvent the Sheng-Suzuki theorem, one must include
the operator $[V,[T,V]]$ in the factorization process. Unfortunately,
it not possible to go beyond fourth-order by including more operators.
It has been shown\cite{chin051} that a
forward sixth-order propagator would have required
the operator $[V,[T,[T,[T,V]]]]$, which is non-separable
and impractical to implement. More recently, by fine-tuning a
family of fourth-order forward algorithm with two
free parameters\cite{chinchen02}
such that the fourth-order error is zero,
Sakkos, J. Casulleras and J. Boronat\cite{casulleras},
and later also one of us\cite{mayrhoferDiplom},
have achieved sixth-order convergence in computing the energy of
a number of quantum systems including liquid $^4$He. Despite this
spectacular advance, it must be noted that the fine-tuning
must be done, in principle, for each individual observable.
The algorithm is therefore only ``quasi-sixth-order'' rather than
uniformly sixth-order.

In this paper, we will present the first QMC simulations using
a bona fide sixth-order and eighth-order algorithm for imaginary time propagation.
The algorithm is based on the multi-product expansion\cite{chin084}
of the short time propagator, which is a new way of achieving higher order
convergence circumventing the Sheng-Suzuki theorem.
This is reviewed in Section~\ref{ssec:multi} 
followed by a brief introduction to path integral ground
state Monte Carlo (PIGSMC) in section~\ref{ssec:PIGSMC}.

\section{Theory}
\label{sec:theory}

\subsection{Multi-product expansion of $G$}
\label{ssec:multi}

Let $G_2(\ep)$ denote the second-order split propagator~(\ref{g2form}),
then for a given set of $n$ whole numbers $\{k_i\}$,
the multi-product expansion of Ref.\onlinecite{chin084} yields the
following second-order propagator:
\begin{equation}
  G_{2n}(\ep) = \sum_{i=1}^{n}  c_i G_2^{k_i}(\ep/k_i)=G(\ep) + O(\ep^{2n+1})
\label{gextra}
\end{equation}
where the expansion coefficient has the closed form
\be
c_i=\prod_{j=1 (\ne i)}^n\frac{k_i^2}{k_i^2-k_j^2}.
\la{coef}
\ee
For PI(GS)MC, it is convenient to choose the sequences
$\{k_i\}=\{1,2\},\{1,2,4\}$ and $\{1,2,3,6\}$ to produce the following
fourth, sixth and eighth-order propagators:
\begin{align}
  G_4(\ep) &=
  -\frac13 G_2(\ep)+\frac43 G_2^2\left(\frac{\ep}2\right)
  \label{eq:G24} \\
  G_6(\ep) &= \frac1{45} G_2(\ep)
                     - \frac49 G_2^2\left(\frac{\ep}2\right)
                   + \frac{64}{45} G_2^4\left(\frac{\ep}4\right)
  \label{eq:G26}\\
  G_8(\ep) &=-\frac1{840} G_2(\ep)
             +\frac2{15} G_2^2\left(\frac{\ep}2\right)\nn\\
&            -\frac{27}{40} G_2^3\left(\frac{\ep}3\right)
             +\frac{54}{35}G_2^6\left(\frac{\ep}6\right).
  \label{eq:G28}
\end{align}
As we will see later, these sequences are chosen because
they are the minimal ``commensurate'' sequences.
Schmidt and Lee\cite{kevin95} have previously suggested the use of (\ref{eq:G24})
in path integrals and did use it in computing the two-particle density
matrix. However, they did not suggest that it can be used for doing path integral
{\it Monte Carlo} simulations.

Since $G(\ep)> 0$, only the error terms in eq.~(\ref{gextra})
can be negative. Thus for sufficiently small $\ep$,
these extrapolated propagators, despite the explicit subtractions,
must be positive.  Only when $\ep$ is sufficiently large, the
error terms overwhelm $G(\ep)$ in a significant fraction of configuration
space.  We will see below that such large $\ep$ cannot be used anyway
because the propagators become highly inaccurate.
One might argue that the error terms can be so singular that despite
the smallness of $\ep$, it can overwhelm $G(\ep)$ at some specific
locations. However, this cannot happen, because by construction
$G_2(\ep)$ is bounded everywhere (except in the case of the Coulomb potential, 
which is a well known
problem\cite{li} in PIMC and which we exclude from
the present consideration).
The subtraction of two bounded functions cannot be singular.
This point will be clear when we present the explicit construction
of extrapolated propagators and 
numerical results in the following sections.

\subsection{Path integral ground state Monte Carlo}
\label{ssec:PIGSMC}

The above multi-product propagators can be applied to any general
PIMC simulations. Here, we will implement it in the specific
context of PIGSMC.
PIGSMC samples the whole probability distribution
function corresponding to a discretized imaginary time propagation
from a trial wave function $\Psi_T(R)$ to the (in principle)
exact ground state $\Psi_0(R)$, where $R$ denotes all degrees of
freedom, e.g. for the translational
coordinates of $N$ particles, $R=(\qr_1,\dots,\qr_N)$.

For any trial wave function $\Psi_T$ with non-zero overlap with the
exact ground state, the exact ground state wave function 
can be obtained by evolving in imaginary time
\begin{equation*}
  \Psi_0(R)\propto
  \lim_{\beta / 2 \rightarrow \infty} \int G(R,R',{\frac{\beta}2 }) \Psi_T(R') dR'.
\end{equation*}
$G({\frac{\beta}2})$ is evaluated by factorizing it into
a product of small time step propagators $G(\ep)$, $\ep=\frac{\beta}{2M}$,
which can be approximate by one of the above-mentioned
short time approximations. Therefore, the full probability distribution to be sampled is
\begin{align*}
    P(R_0,\dots,R_{2M}) =&  {\frac1{\cal N}} \Psi_T(R_0)\
      G(R_0,R_1;\ep)\dots\\
      &G(R_{2M-1},R_{2M};\ep)\ \Psi_T(R_{2M}),
\end{align*}
so that the expectation value $\langle\Psi_0| A|\Psi_0 \rangle$
of a local operator $A(R)$ is evalutated by sampling $A$
at the central time step, $A(R_M)$.
For the energy, we take advantage of $[H,G]=0$ to obtain the
energy estimator in terms of the local energy of the trial
wave function $E_L(R)={H\Psi_T/\Psi_T}$
\begin{equation*}
  E_0 = \int\! dR_0\dots dR_{2M}\ E_L(R_0)\, P(R_0,\dots,R_{2M})\,.
\la{hexp}
\end{equation*}
These multidimensional integrations can be carried out
with the Metropolis method.

\subsection{Implementing multi-product expansions in PIGSMC}
\label{ssec:implement}

To see how one can implement these multi-product propagators in PIGSMC,
we will now give a detailed discussion of the fourth order case. 
Considering $G_4$ at time step size $2\epp$:
\begin{align*}
 G_4(2\epp)&=\frac43\e^{-\epp V/2}\e^{-\epp T}\e^{-\epp V}\e^{-\epp T}\e^{-\epp V/2}\\
        &-\frac13\e^{-\epp V}\e^{-2\epp T}\e^{-\epp V}.
\end{align*}
In evaluating the matrix element of $G_4(2\epp)$, since the first term on the RHS
has one more operator $\e^{-\epp T}$, it would require
one more intermediate state integration than the second term,
resulting in two dissimilar terms difficult to sample uniformly.
The key contribution of this work is to enforce uniformity by
artificially splitting the single operator $\e^{-2\epp T}$ in the second term into two:
\begin{align}
 G_4(2\epp)&=\frac43\e^{-\epp V/2}\e^{-\epp T}\e^{-\epp V}\e^{-\epp T}\e^{-\epp V/2}\nonumber\\
        &-\frac13\e^{-\epp V}\e^{-\epp T}\e^{-\epp T}\e^{-\epp V},
\label{eq:G24rA}
\end{align}
which gives in the coordinate respresentation
\begin{align}
&  \langle 1| G_4(2\epp)|3\rangle
 =\int d2 \langle 1|\e^{-\epp T}|2\rangle \langle 2|\e^{-\epp T}|3\rangle\nonumber\\
&\qquad\qquad \times \Big[\frac43 \e^{-\epp V_1/2-\epp V_2-\epp V_3/2}
  -\frac13 \e^{-\epp V_1-\epp V_3}\Big] \nonumber\\
&=\int d2 G_0(12;\epp)G_0(23;\epp)\e^{-\epp V_1/2-\epp V_2-\epp V_3/2}F(123,\epp)
\label{g4pimc}
\end{align}
where we have denoted $V_k=V(R_k)$ and abbreviated $R_k\rightarrow k$.
We have defined
\be
 F(123,\epp)=\frac13 \Big[4- \e^{-\epp \left(\frac{V_1+V_3}2 - V_2\right)}\Big],
\la{ffct}
\ee
and the free propagator
\be
G_0(12;\epp)=\langle 1|\e^{-\epp T}|2\rangle
=(4\pi D\epp)^{-3N/2} \e^{- \frac{(R_1 - R_2)^2}{4D\epp} }.
\ee
We observe that: ({\it i\/}) Without the factor $F$, (\ref{g4pimc}) is just accurate to
second order. ({\it ii\/}) By including $F$, only the potential energy is extrapolated
in order to convert $G$ to fourth-order.
({\it iii\/}) For sufficiently small $\ep$, $F>0$. ({\it iv\/}) If the potential
function is mostly convex (such as Lennard-Jones type potential near the
potential minimum), then one has
\be
  \frac{V(R_1)+V(R_3)}2\ge V\Big(\frac{R_1+R_3}2\Big).
\la{convex}
\ee
Since
\begin{equation*}
  G_0(12;\epp)G_0(23;\epp)=
  \frac{\e^{-\frac1{2D\epp} (R_2 -\frac{R_1+R_3}2)^2}}{(2\pi D\epp)^{3N/2}}
  G_0(13;2\epp),
\end{equation*}
for fixed $R_1$ and $R_3$, $R_2$ is normally distributed
about $(R_1+R_3)/2$ with width $\propto \sqrt{\epp}$. If $R_2$ is such that it is between
$R_1$ and $R_3$, then the convexity condition (\ref{convex}) would guarrantee (\ref{ffct})
to be positive for all $\epp$. This only fails when the width of the Gaussian distribution for
$R_2$ exceeds $|R_1-R_3|/2$, suggesting that the near-positivity
of $F$ can extend over a rather wide range of $\epp$, which is indeed observed.
Metropolis sampling requires exact positivity of $F$, which we ensure by
using $\max(0,F)$, i.e. rejecting moves where $F<0$.
We also collect statistics about these rejections, so ensure
that their rate is low, and decreasing with $\epp$.

The generalization to higher order is now clear. For the sixth-order, Eq.(\ref{eq:G26}),
\begin{align}
  G_6(4\epp) &=
  \frac{64}{45}\e^{-\epp V/2}\e^{-\epp T}\e^{-\epp V}\e^{-\epp T}\e^{-\epp V}\e^{-\epp T}\e^{-\epp V}\e^{-\epp T}\e^{-\epp V/2}\nn\\
&-\frac{4}{9}\e^{-\epp V}\e^{-\epp T}\e^{-\epp T}\e^{-2\epp V}\e^{-\epp T}e^{-\epp T}\e^{-\epp V}\nn\\
&+ \frac1{45}\e^{-2\epp V}\e^{-\epp T}\e^{-\epp T}\e^{-\epp T}\e^{-\epp T}\e^{-2\epp V},
\end{align}
yielding the coordinate representation
\begin{align}
  G_6(12345;4\epp) &=
  G_0(12;\epp) G_0(23;\epp) G_0(34;\epp) G_0(45;\epp) \nn\\
& \times \Big[\frac{64}{45} \e^{-\epp V_1/2-\epp V_2-\epp V_3-\epp V_4-\epp V_5/2}\nn\\
&   -\frac49 \e^{-\epp V_1-2\epp V_3-\epp V_5} + \frac1{45}\e^{-2\epp V_1-2\epp V_5} \Big].
\la{g6pimc}
\end{align}
Similarly for the eighth-order (\ref{eq:G28}),
\begin{align}
& G_8(1234567;6\epp) =
  G_0(12;\epp)G_0(23;\epp)\dots G_0(67;\epp)\nn\\
& \times\Bigl[ \frac{54}{35}\e^{-\epp V_1/2} \e^{-\epp V_2}\e^{-\epp V_3}\e^{-\epp V_4}\e^{-\epp V_5}\e^{-\epp V_6}\e^{-\epp V_7/2} \nn\\
& -\frac{27}{40}\e^{-\epp V_1} \e^{-2\epp V_3}\e^{-2\epp V_5}\e^{-\epp V_7}\nn\\
& + \frac2{15}\e^{-3\epp V_1/2}\e^{-3\epp V_4}\e^{-3\epp V_7/2}
  - \frac1{840}\e^{-3\epp V_1}\e^{-3\epp V_7} \Bigr]\,.
\la{g8pimc}
\end{align}
For commensurate sequences one
can factor out all the free-propagators and restrict the extrapolation process 
only to the potential energy function.

\section{Results}

We have implemented the PIGSMC algorithm using multi-level sampling
as described in the review~\cite{ceperleyRMP95}.  We compare
our new extrapolated fourth, sixth, and eighth order propagators with the primitive (second-order)
and the fourth-order forward propagators\cite{chin97} 4A
\begin{equation}
  G_{4A}(\ep) =
  \e^{-{\ep\over 6}V}
  \e^{-{\ep\over 2}T}
  \e^{-{2\ep\over 3}V-\frac{\ep^3}{72}[V,[T,V]]}
  \e^{-{\ep\over 2}T}
  \e^{-{\ep\over 6}V}\,.
\label{eq:Gcomm}
\end{equation}

\begin{figure}[hbt]
\includegraphics[width=0.8\linewidth]{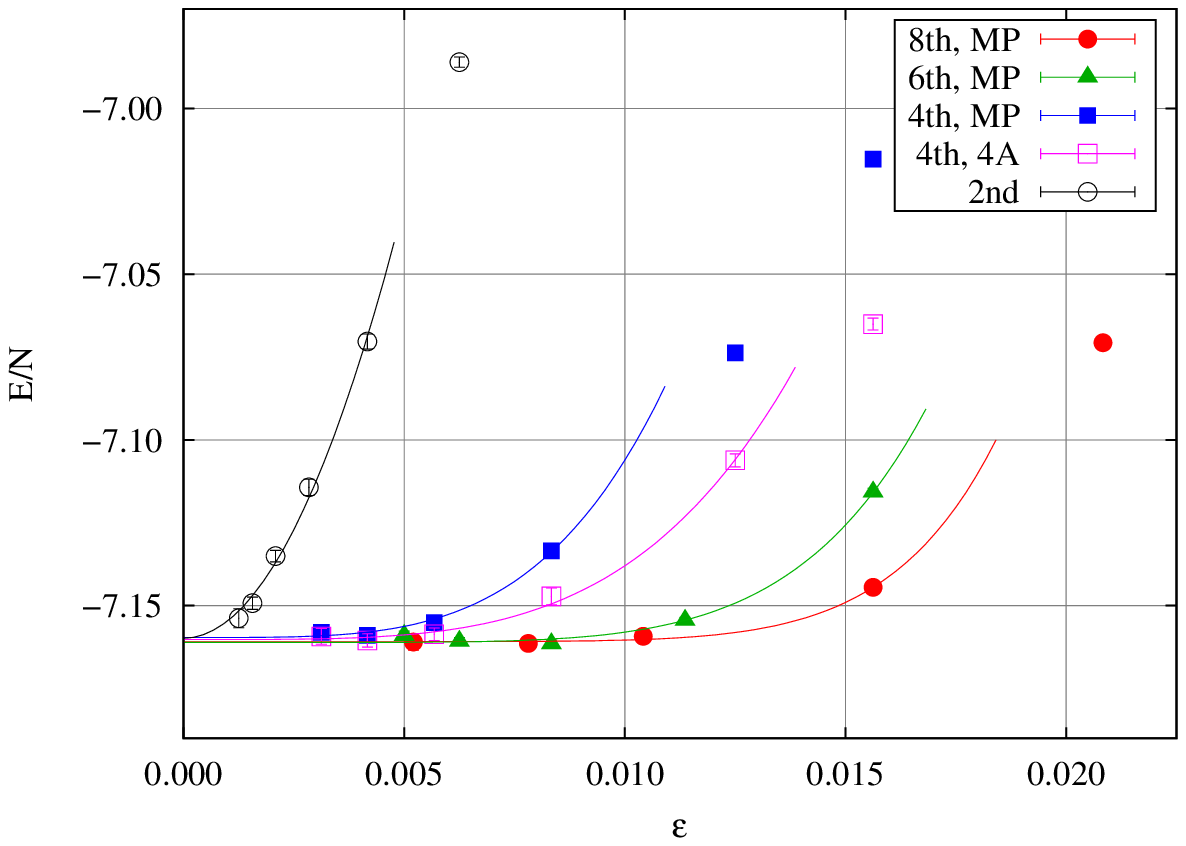}
\caption[]{\label{FIG:E} (color online) Ground state energy $E_0$ of bulk $^4$He,
  simulated by 64 $^4$He atoms, as a function of imaginary time step
  $\ep$.  Decay time was $\beta=0.25$K$^{-1}$.  We compare results produced
  by the primitive second-order 
  propagator $G_2(\ep)$ and the fourth-order forward propagator\cite{chin97} $G_{4A}(\ep)$
  (denoted ``4A'')
  with our fourth, sixth, and eighth-order multi-product propagators $G_4(\ep)$,
  $G_6(\ep)$ and $G_8(\ep)$,
  (denoted ``MP'').
  Each $E_0(\ep)$ is fitted with the appropriate polynomial.
}
\la{fig1}
\end{figure}

To demonstrate that our multi-product propagators work for realistic,
and strongly interacting quantum systems, we apply them to the case of bulk liquid $^4$He.
We calculate the ground state energy $E_0$
at equilibrium density $\rho_0=0.02186$\AA$^{-3}$, by a PIGSMC simulation
of 64 $^4$He atoms in a simulation box with periodic boundary conditions.
The decay time is $\beta=0.25$K$^{-1}$, and we use the potential by
Aziz et al.~\cite{aziz79}.  In Fig.~\ref{FIG:E} we show $E_0/N$ as function
of $\ep$ for various propagators.  We fit the polynomial
$a+b\ep^n$ (lines) to $E_0(\ep)/N$, where $n$ is the order
of the respective propagator.  Since the order of $E_0(\ep)$ is defined as the
$\ep\to 0$ behavior, we have restricted the fits to small values of $\ep$ --
the end point of the lines indicate the fitting interval.
These propagators
are compared at equal time steps: $G_2(\ep)$, $G_4(\ep)$, $G_6(\ep)$,
and $G_8(\ep)$, to verify the order of convergence.

The primitve second-order propagator (open circle)
is clearly a poor approximation, with a large error even
for small $\ep$, and therefore requires a large number of
beads.  The simplest fourth-order forward propagator 4A, Eq.(\ref{eq:Gcomm}),
is a significant improvement, as can be seen in the 
behavior of $E_0(\ep)/N$ (open square), with error coefficient
smaller than our fourth-order multi-product propagator~(\ref{g4pimc})
(filled square).  However, the forward 4A propagator requires the 
the computation of $[V,[T,V]]\propto |\nabla_i V|^2$, and
its relative efficiency would depend on the complexity
of evaluating this gradient.  Both can be fitted well by a 
fourth-order polynomial with $n=4$.  Finally, the closed triangle and circle
show the convergence of the sixth order~(\ref{g6pimc}) and
eighth order~(\ref{g8pimc}) multi-product expansion, which indeed has a smaller
$\ep$ dependence in the range of Fig.~\ref{FIG:E}.
These multi-product propagators are true high order propagators and
will produce sixth and eighth order convergence for the expectation value
of any observable. The present results constitute the first
implementation of a quantum Monte Carlo simulation with
a bona fide imaginary time propagator of higher than fourth order.
At small values of the time step size, say at $\ep=0.005$, the
sixth and eighth order algorithms produce very precise results which
are not indicated by Fig.\ref{fig1}.

The multi-product expansion of $G$, Eq.~(\ref{gextra}), is not strictly positive everywhere
for a finite time step $\ep$, as we have discussed above for
the fourth-order case.
In Fig.~\ref{figx} we show the ratio $R_n$ of
attempted MC moves that are rejected due to negativity of the multiproduct
expansion.  $R_n$ is decreasing with $\ep$ as expected.  Only in the
sixth-order case that we observed a non-monotonous behavior at
very large $\ep$, that the ratio $R_n$ {\em decreases} with increasing $\ep$.
This occurs at large values of $\ep$ where 
the error of the energy $E_0(\ep)/N$ is rapidly
increasing with $\ep$ (outside the range of Fig.~\ref{fig1}).
This may due to system configurations with complicated intertwining
negative and positive regions of $G$.
We want to stress that, since this happens only
for $\ep$ much too large for quantitatively correct results, it
poses no practical limitations.

The convergence plot Fig.~\ref{FIG:E} does not reveal the actual
computational effort required for a desired accuracy.
From our derivation, it is clear that the computational effort of
$G_4$, $G_6$, and $G_8$ are roughly
equivalent to running $G_2$ twice, four, and six times, respectively.
This then means that for a given $\ep$ for $G_2(\ep)$, one should
compare it to $G_4$ at $2\ep$, $G_6$ at $4\ep$ and $G_8$ at $6\ep$,
that is, for an equal effort comparison, we should compare
$G_2(\ep)$, $G_4(2\ep)$, $G_6(4\ep)$ and $G_8(6\ep)$.
This is done in Fig.~\ref{FIG:Eeffort}.
In this comparison, at a given $\ep$ each algorithm uses the same
number of beads. For the forward algorithm 4A, this comparison
neglects the additional cost of evaluating the gradient $|\nabla_i V|^2$.
If $|\nabla_i V|^2$ required no more effort than that of
evaluating the potential, then
propagator 4A would actually outperform the extrapolated
sixth- and eighth-order algorithms at time steps $\ep\gtrsim 0.003$K$^{-1}$.
This confirms that, in principle, a purely forward time step algorithm
can be more efficient, provide that $|\nabla_i V|^2$ can be easily evaluated.
However, the higher order extrapolated algorithms are clearly easier to
derive and implement. Morever, when very high accuracy
is required, such the ``chemical accuracy'' required in quantum chemistry
applications, then a higher order algorithm will always outperform a lower
order algorithm.  This is specially critical in determining the equilibrium
configuration or conformation of clusters and macromolecules, where
energy differences are very small.

\section{Conclusion and Outlook}

In this work, we have shown how to implement the multi-product expansion
of the imaginary time propagator in QMC for solving strongly
interacting quantum many-body systems, such $^4$He,
to any desired order in the imaginary time step $\ep$.
In particular this work is the first demonstration of
truly sixth- and eighth-order QMC algorithms.  In the case of
$^4$He, our results suggest that these higher than fourth-order
algorithms may not be more efficient than purely forward time
step fourth-order algorithms, but they do have the simplicity of
not requiring the potential gradient. This is particularly
useful in simulating non-cartesian
coordinate systems, such as molecules~\cite{zillichJCP05}
with anisotropic constituents and rotational degrees of freedom.
Moreover, these extrapolated propagators are the only higher order
algorithms possible in cases where the double-commutator cannot be
evaluated, such as for the diatoms-in-molecule potential~\cite{leinoJCP08}.
Finally, for QMC applications where chemical accuracy is required,
such as in determining equilibrium configurations and conformations,
our sixth and higher order multi-product propagators
will be computationally more efficient than fourth-order propagators.

\begin{acknowledgments}
We are grateful for helpful discussions with Eckhard Krotscheck.
The work was supported by the Austrian Science Fund FWF, project number
P21924.
\end{acknowledgments}

\newpage

\newpage

\begin{figure}[hbt]
\includegraphics[width=0.8\linewidth]{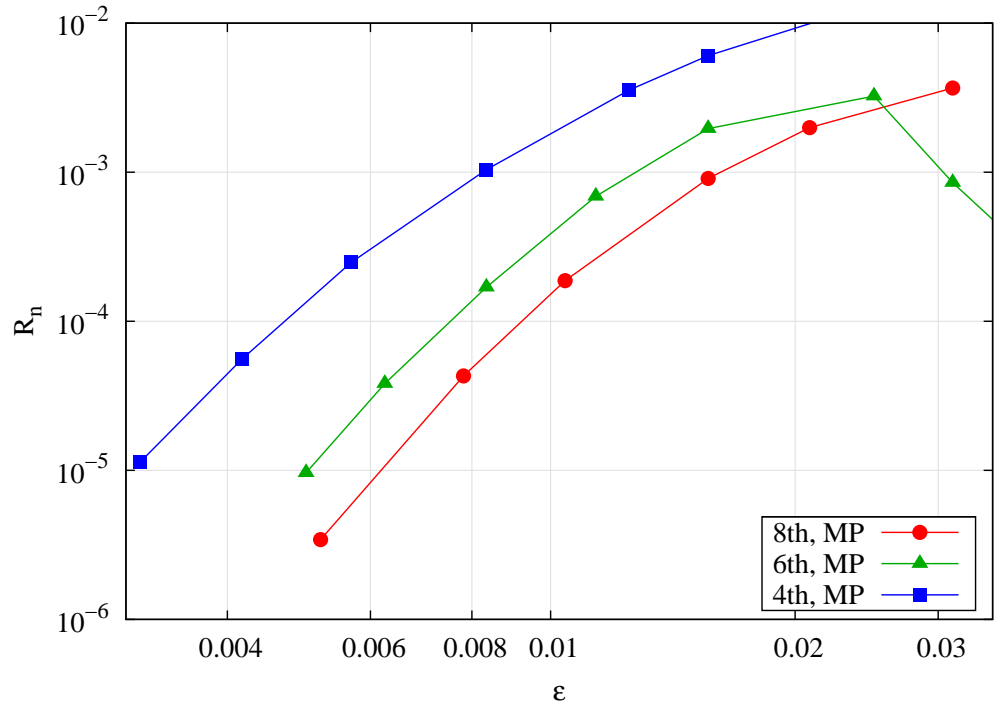}
\caption[]{(color online)
  The ratio $R_n$ of rejected MC moves that would lead to a negative
  propagator $G$.  $R_n$ is decreasing with time step $\ep$.
}
\la{figx}
\end{figure}
\newpage

\begin{figure}[hbt]
\includegraphics[width=0.8\linewidth]{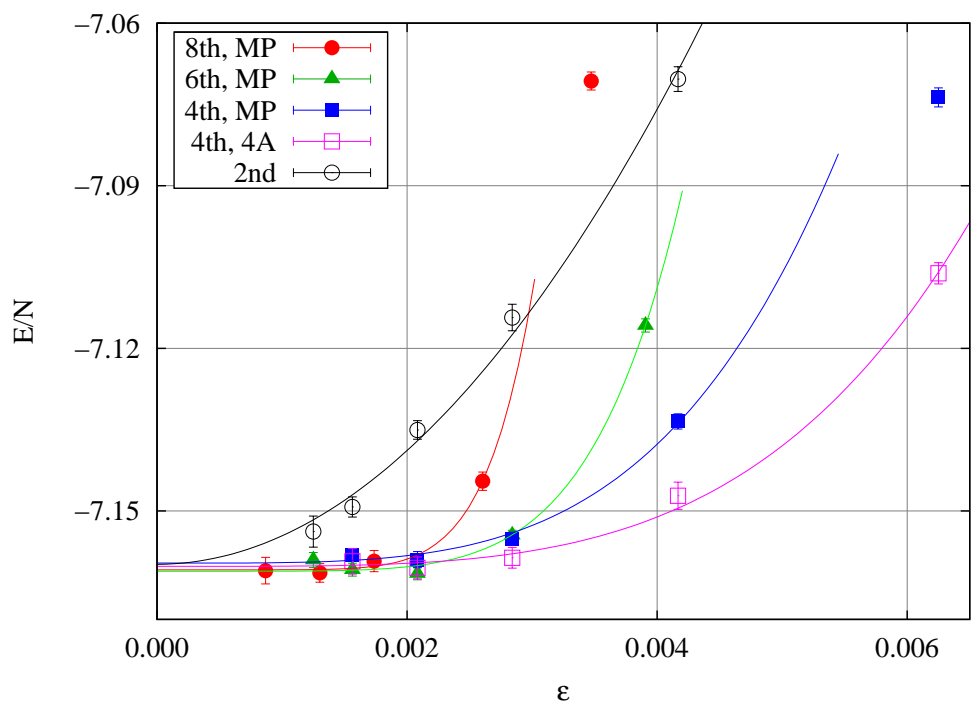}
\caption[]{\label{FIG:Eeffort} (color online)
A roughly equal effort comparison of algorithms $G_2(\ep)$,
$G_4(2\ep)$, $G_6(4\ep)$, $G_8(6\ep)$ and $G_{4A}(2\ep)$
for the same ground state energy $E_0$ as
  in Fig.~\ref{FIG:E}.
}
\la{fig2}
\end{figure}

\end{document}